\documentclass[a4paper, amsfonts, amssymb, amsmath, reprint,reprintnumbers, showkeys, nofootinbib, twoside, floatfix]{revtex4-1}
\usepackage{float}
\usepackage{dsfont}
\usepackage[english]{babel}
\usepackage[utf8]{inputenc}
\usepackage[colorinlistoftodos, color=green!40, prependcaption]{todonotes}
\usepackage[pdftex, pdftitle={Article}, pdfauthor={Author}]{hyperref} 
\usepackage{graphicx}
\usepackage{epsfig,rotate,latexsym,amsmath,mathrsfs,amssymb}
\usepackage{xcolor,hyperref}
\usepackage[arrowdel]{physics}
\hypersetup{final}
\def \be{\begin{equation}}
\def \ee{\end{equation}}

\def \nn{\nonumber}

\begin{document}
\preprint{IMSC/2025/06}

\keywords{acoustic black hole, Hawking radiation, ultra-relativistic 
heavy-ion collisions} 

\title{Doppler shifted Hawking radiation from acoustic black holes in 
ultra-relativistic heavy-ion collisions}

\author{Sanatan Digal}
\email{digal@imsc.res.in}
\affiliation{Institute of Mathematical Sciences, Chennai 600113, India}
\affiliation{Homi Bhabha National Institute, Training School Complex, Anushaktinagar, Mumbai 400094, India}
\author{Ajit M. Srivastava}
\email{ajit@iopb.res.in}
\affiliation{Institute of Physics, Bhubaneswar 751005, India}

\begin{abstract}
In a hydrodynamic flow, with flow becoming supersonic at some point,
the subsonic-supersonic boundary behaves as the horizon of a
black hole. Possibility of detecting Hawking radiation from such 
{\it acoustic black holes} has been investigated in 
a variety of laboratory systems, ranging from cold atom systems, to
condensed matter systems with hydrodynamic flow of electrons, to
relativistic heavy-ion collisions (at relatively lower collision energies).
Ultra-relativistic heavy-ion collisions, with boost-invariant longitudinal
flow of the quark-gluon plasma (QGP) in a wide rapidity window has eluded 
this remarkable possibility because in this case the black hole 
horizon is dynamical, moving away from
center with sound velocity, leading to infinite red shift of Hawking 
radiation. We show here that such a conclusion is premature. The QGP 
flow at very large rapidities, necessarily
deviates from Bjroken boost invariant flow. Due to this, an observer
close to that region sees black hole horizon with a finite
redshift. It leads to non-trivial prediction of Hawking radiation affecting
particle momentum distributions for a window of rapidities, leaving
near central rapidity regions unaffected.
\end{abstract}

\pacs{PACS numbers: 04.70.Dy, 04.80.Cc, 12.38.Mh}

\maketitle

\section{Introduction} \label{sec:intro}
Analogue gravity has opened new directions of investigating certain 
aspects of General Relativity in laboratory systems 
\cite{Unruh:1980cg,Visser:1997ux}.  One of the most exciting  possibility is 
the so called acoustic black holes, in particular the physics of Hawking 
radiation.
In fact, most of the investigations of acoustic black holes have focused
on the possibility of observing Hawking radiation in various laboratory
systems. By analyzing the propagation of acoustic perturbations in the velocity
potential of an inviscid, irrotational, 
barotropic fluid, Unruh had shown \cite{Unruh:1980cg}
that these acoustic perturbations obey an equation which is same as the Klein
Gordon equation for a massless scalar field in a curved Lorentzian spacetime,
with the spacetime metric determined by the flow  velocity, density, and 
pressure of the fluid. If the flow changes from subsonic to supersonic,
then the surface on which the normal component of fluid velocity equals
the local speed of sound, becoming supersonic beyond it, behaves as the 
horizon of a black hole, as acoustic perturbations  cannot cross this surface 
from the supersonic region to the subsonic region. This intuitive expectations
was shown to be rigorously true by Unruh \cite{Unruh:1980cg} who showed
that for  a spherically symmetric, stationary, convergent background fluid 
flow, the effective metric seen by acoustic perturbations of the velocity 
potential is  the Schwarzschild metric, with the black hole horizon
coinciding with the surface where the fluid velocity becomes supersonic.
It then follows \cite{Unruh:1980cg} that for the case when acoustic 
perturbations can be quantised, a thermal bath of acoustic phonons will be 
expected to be emitted from this sonic horizon in the form of Hawking radiation. 

A variety of laboratory systems have been proposed where this possibility 
of acoustic black holes may be realized and resulting Hawking radiation may
be detected \cite{Steinhauer:2015ava,Steinhauer:2015saa,MunozdeNova:2018fxv,
Wang:2016jaj,Michel:2016tog,Grisins:2016gru,
Liberati:2019fse,Isoard:2019buh,Jacquet:2021scv,Syu:2022ajm,
Tian:2018srd,Ganguly:2019mza,Kolobov:2019qfs,Barcelo:2005fc}.  
First experimental realization was with cold atom systems where
the observation of Hawking radiation in terms of correlated pairs of
Hawking particles emitted from the sonic horizon has been claimed with
the two partners of the pair propagating on the two sides of the sonic horizon
\cite{Steinhauer:2015ava,Steinhauer:2015saa,MunozdeNova:2018fxv}, (see, also
\cite{Wang:2016jaj,Michel:2016tog,Grisins:2016gru,Liberati:2019fse,
Isoard:2019buh,Jacquet:2021scv,Syu:2022ajm} in this context).
Some of us recently investigated this 
possibility \cite{Dave:2022wgk} for certain condensed matter systems where 
it has now become possible to realize the hydrodynamic regime for electron 
transport. It was shown in ref. \cite{Dave:2022wgk} that
for a 2-dimensional sample with the converging-diverging geometry
of a de Laval nozzle, it is possible to realize electron flow pattern
suitable for acoustic black hole, with the black hole horizon forming at
the neck of the de Laval geometry. Resulting Hawking radiation in
this case should then be observable in terms of current fluctuations. It
was further argued that current fluctuations on both sides of the acoustic 
horizon should show correlations expected for pairs of Hawking particle.

Apart from these very low energy/temperature condensed matter systems for
acoustic black holes, a completely different regime of very high 
energy/temperature was proposed by some of us \cite{Das:2020zah} for 
relativistic 
heavy-ion collisions. It was shown in \cite{Das:2020zah} that an acoustic black 
hole metric may be constructed in the flow of quark-gluon plasma (QGP) in 
relativistic heavy-ion collisions \cite{Das:2020zah}, with the resulting thermal
radiation of acoustic phonons observable in terms of modification of
the rapidity dependence of the transverse momentum distribution of
various particles. One limitation of the analysis in \cite{Das:2020zah} was
that it was applicable for relatively lower energy relativistic heavy-ion
collisions with resulting plasma temperature of order 135 MeV for which it
was shown that resulting Hawking temperature will be about 4-5 MeV.

It is important to note that 5 MeV of Hawking temperature, with background
plasma having temperature of 135 MeV does not mean at all that the signal of
Hawking radiation will be highly suppressed. This is because these two
temperatures refer to completely different physical quantities. While plasma 
temperature directly relates to the $p_T$ distribution of emitted particles,
the Hawking radiation temperature corresponds to the temperature of phonon 
system which represents fluctuations in the longitudinal flow velocity
(potential). Low value of this temperature for the 
phonon system corresponds to longer wavelengths, 
meaning that $p_T$ distribution of relatively widely separated rapidity
windows  will get mixed up. A detailed analysis of the resulting signal
in terms of perturbations on the $p_T$ distribution is underway and we
hope to report it in a future work.

  The reason that the analysis in \cite{Das:2020zah} was restricted to lower
energy collisions was due to expected boost-invariance of the longitudinal
flow of the quark-gluon plasma (QGP) in a wide rapidity window 
in ultra-relativistic heavy-ion collisions at very high energies, such 
as at LHC and the high center of mass
energy collisions at RHIC.  In that case, the black hole horizon is dynamical, 
moving away from center with sound velocity, which leads to infinite red 
shift of Hawking radiation (see, refs. 
\cite{Ganguly:2019mza,Nielsen:2008cr,Nielsen:2005af} in this context).
 At low energy collisions, the longitudinal flow
does not follow Bjorken boost invariance, except possibly in a very
narrow central rapidity region. Energy density gradients lead to
acceleration of flow which can (almost) compensate for the slow down
of the flow due to expansion. using URQMD simulations, it was shown
in \cite{Das:2020zah} that for a limited duration of proper time, it is then 
possible to achieve static horizon with observable predictions for
the associated Hawking radiation.

In this paper, we will reconsider this chain of arguments and show that 
even at ultra-high energy collisions, e.g. the highest center of mass 
collisions at LHC with very wide range of rapidity with boost invariant 
longitudinal flow for considerable duration of time, it is possible to 
have observable signals of Hawking radiation. The QGP flow velocity
in this regime has only z-component $v^z$ (along the beam axis), with 
the scaling law $v^z = z/t$ as measured in the centre of mass frame. 
$t = 0$ corresponds to the instant when the two (highly Lorentz contracted) 
nuclei overlap. The velocity field $\vec{v}(t, z) = (0,0, v^z (t,z))$ is
naturally irrotational, with which one can express it in terms of a scalar 
velocity potential $\psi (t, z)$ as $\vec{v} = \vec{\nabla} \psi$. With
this, and with barotropic equation of state (see, ref.\cite{Das:2020zah} 
for details),
the plasma expansion satisfies all the criteria necessary for the 
construction of an analogue model of gravity. We directly write down the 
effective acoustic metric for this system in coordinates $(t,x,y, z)$ 
\cite{Visser:1997ux,Ganguly:2019mza}:
\begin{align}
ds^2
 =
\frac{\rho(t, z)}{ c_s(t, z)}
\Big[
& \, - \left(  c_s(t, z)^2- v^{ z}(t, z)^2 \right) dt^2
 - 2  v^{ z}(t, z) dt d z      \nn \\
& \, + dx^2 + dy^2 + d z^2
\Big]~.
\label{eq:acmet}
\end{align}
Here, $ c_s(t, z)$ is the speed at which acoustic perturbations
propagate in the fluid with  $c_s^2=\partial {p}/\partial \rho$ with $p$
and $\rho$ being the pressure and the energy density respectively. 
If the velocity field of the fluid is such that, for some value $z=z_H$,
\begin{align*}
v^z & \, < c_s \qquad \text{for} \ z < z_H~, \\
v^z & \, = c_s \qquad \text{for} \ z = z_H~, \\
v^z & \, > c_s \qquad \text{for} \ z > z_H~,
\end{align*}
then an acoustic horizon forms at $ z = z_H$. The fluid flowing with 
supersonic velocities in $z > z_H$ sweeps away all acoustic perturbations 
away from the horizon. The supersonic region is thus acoustically 
disconnected from the subsonic region. As $v^z 
\rightarrow 0$ when $z \rightarrow 0$, we get back Minkowski metric there. Thus, an observer 
at $z = 0$ would serve as an ``asymptotic observer'' in ``asymptotically 
flat'' spacetime for our purposes. With the above metric seen by the 
perturbations in the velocity potential $\psi$, there would be a spontaneous 
emission of phonons (quantised acoustic perturbations) near the horizon in 
the form of \textit{acoustic Hawking radiation}. This radiation should be
thermal, with a temperature given by (in natural units, see, 
ref.\cite{Das:2020zah} for details), 

\begin{equation}
T =  \frac{\kappa}{2\pi} = \frac{1}{2\pi} \frac{\partial v^z}{\partial z} 
\bigg |_{z_H}~.
\label{eq:ht}
\end{equation}
Here, $\kappa$ is known as the surface gravity at the acoustic horizon. 
(The overall conformal factor in the metric in Eqn.\ref{eq:acmet}, with appropriate
asymptotic behavior, does not affect the 
value of the temperature here, see \cite{Das:2020zah}.)

 It is straightforward to see that with $v^z = z/t$ for the Bjorken 
longitudinal scaling expansion, the acoustic horizon (the location
where $v^z = c_s$) is not static, rather it moves away from the centre 
of  collision at the speed of sound. 
We assume constant value for $c_s (= 
1/\sqrt{3})$, which is a reasonable assumption for early times in the central 
rapidity region at such high energy collisions. (Of course one can consider 
changing speed of sound which opens up further possibilities to be considered.
For various condensed matter systems, varying  sound of sound plays important
role in such investigations). Speed of sound playing the role of speed of
light in Eqn.\ref{eq:acmet}, resulting Hawking radiation is infinitely red-shifted
making it unobservable. This situations is schematically represented
by the plot of energy density $\epsilon$ vs. rapidity $\eta$ in Fig.1.

\begin{figure}
\centering
\includegraphics[width=0.85\linewidth]{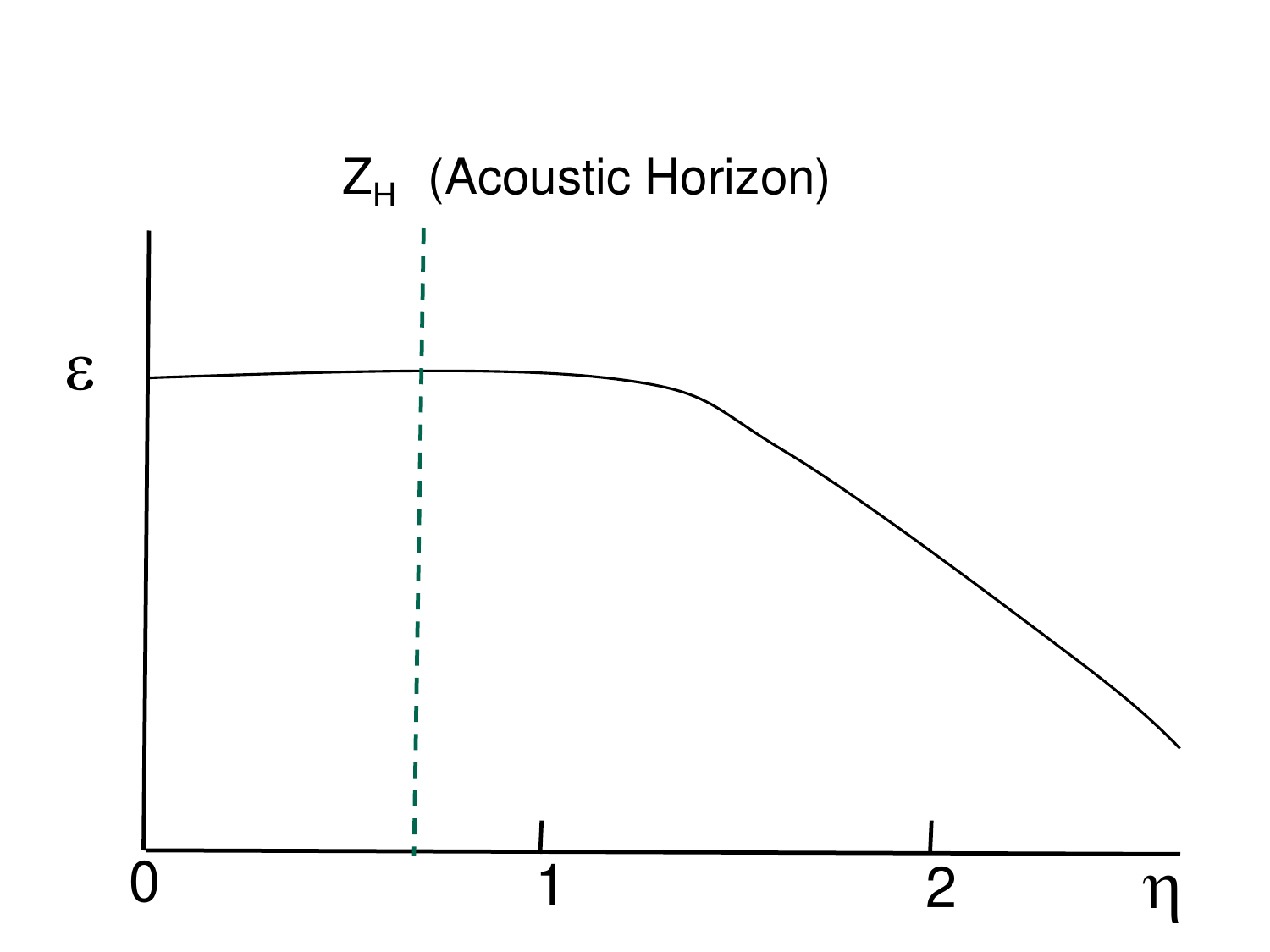}
\caption{\label{fig:fig1} Schematic plot of energy density $\epsilon$ vs. 
rapidity $\eta$. In all the figures, the values on the $\epsilon$ and
$\eta$ axes are arbitrary, only representing qualitative aspect of the 
location of the acoustic horizon within the rapidity plateau region.}
\end{figure}

 This situation was avoided in ref.\cite{Das:2020zah} by considering low energy
 collisions where flat rapidity region was (almost) absent. Energy density
 gradient then leads to fluid acceleration which compensates for the slow
 down of fluid due to expansion, thereby leading to the possibility of
 static horizon. For the case with a flat central rapidity region,
 it is clear from Fig.1, that when acoustic horizon location
 ($z_H$) is deep inside the flat rapidity region, no fluid acceleration is
 possible, and infinite red-shift of Hawking radiation is unavoidable.

  However, as we will argue below, this conclusion is premature. The important
  point to focus is on the notion of observer. So far we have been considering
  the natural choice of observer located at $z = 0$, which represents
  asymptotically flat region for metric in Eqn.\ref{eq:acmet}. Hawking radiation in terms
  of fluid velocity perturbations reaching $z = 0$ region (with phonon
  temperature given by Eq.\ref{eq:ht}), will mix neighboring rapidity 
  distributions of
  transverse momenta $p_T$ of particles, which will be detected by various
  particle detectors.

  However, detectors receive particles coming from different rapidity windows,
  not only from the central rapidity region. Consider particles coming from
  a rapidity window centered at a non-zero value of $\eta = \eta_0$ as 
  shown in Fig.2. An observer comoving with fluid at that rapidity will 
  find its own acoustic horizon located at a different location $z_{H0}$, 
  which, for sufficiently large value of $\eta_0$, may lie in the region of 
  plasma with large energy density gradients, (near, or inside the 
  fragmentation region). For this observer, the fluid expansion in the 
  region near
  $z_{H0}$ will experience acceleration due to energy density gradient,
  (along with the usual slow down due to plasma expansion). This will reduce
  the speed of recession of acoustic horizon with respect to the observer at
  $\eta_0$. If the recession velocity is significantly reduced from the value
  $c_s$, it may make the red-shift of the Hawking radiation finite,
  making it observable.

\begin{figure}
\centering
\includegraphics[width=0.85\linewidth]{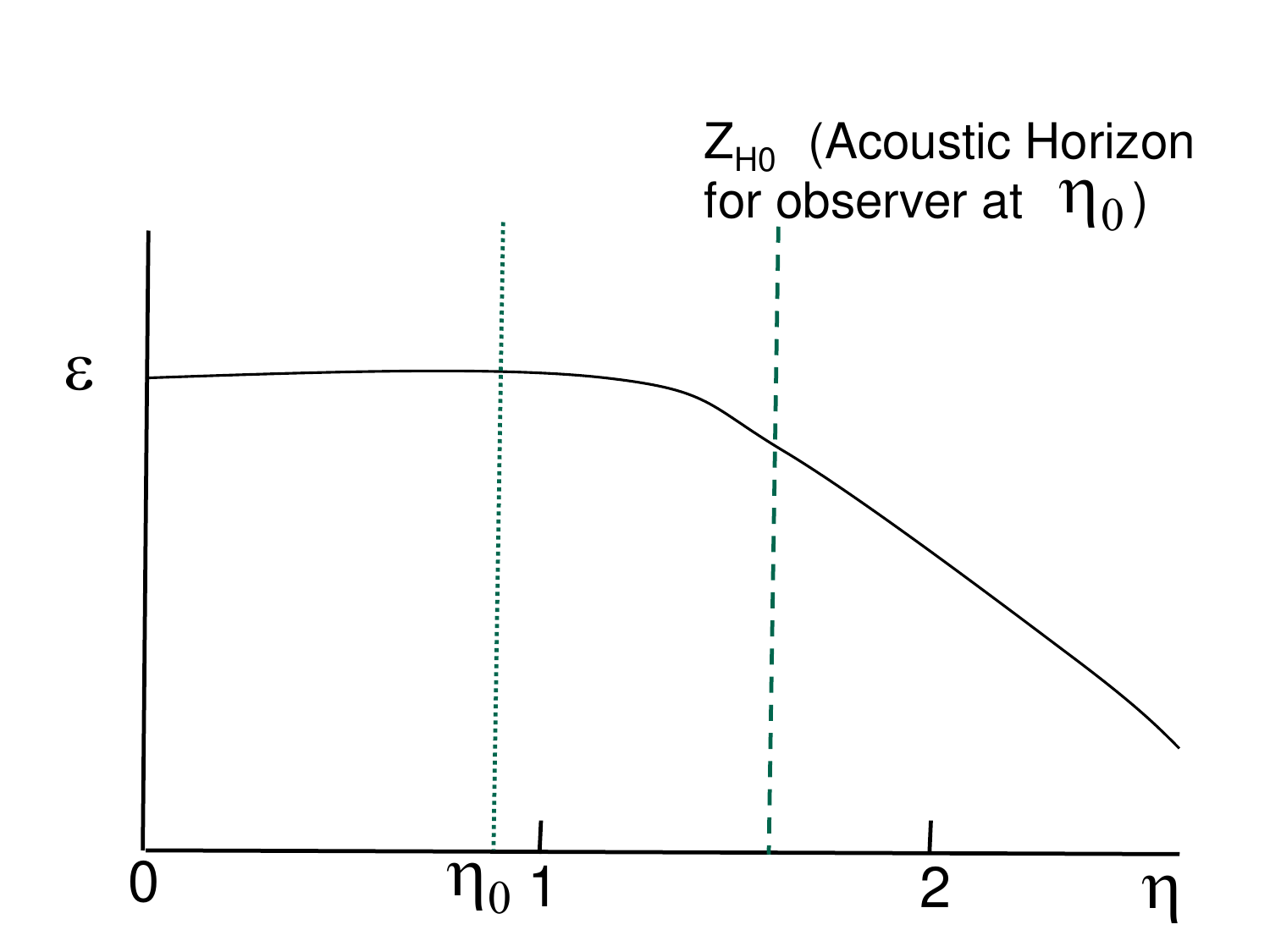}
\caption{\label{fig:fig2} Observer at rest with fluid at non-zero  value of 
$\eta = \eta_0$ such that, the acoustic horizon for this observer lies 
at a different location $z_{H0}$, in the region with non-zero gradient of 
energy density.}
\end{figure}

  Of course, one can hope for parameter space where
  the horizon may be static, or even with blue shift (as was the case for
  certain time duration in \cite{Das:2020zah}). However, our investigation so far
  have not led to such optimistic situation. Still, we are able to find
  significantly reduced recession speeds for the acoustic horizon, with
  red-shift factors for the Hawking radiation of about 2-3. Again, note that
the observer at $\eta_0$ is the plasma at that location where local $p_T$
  distributions of particles will get affected by fluid mixing with 
  neighboring rapidities. This leads to a very interesting pattern of
  effects of Hawking radiation, as shown in Fig.3.

\begin{figure}
\centering
\includegraphics[width=0.85\linewidth]{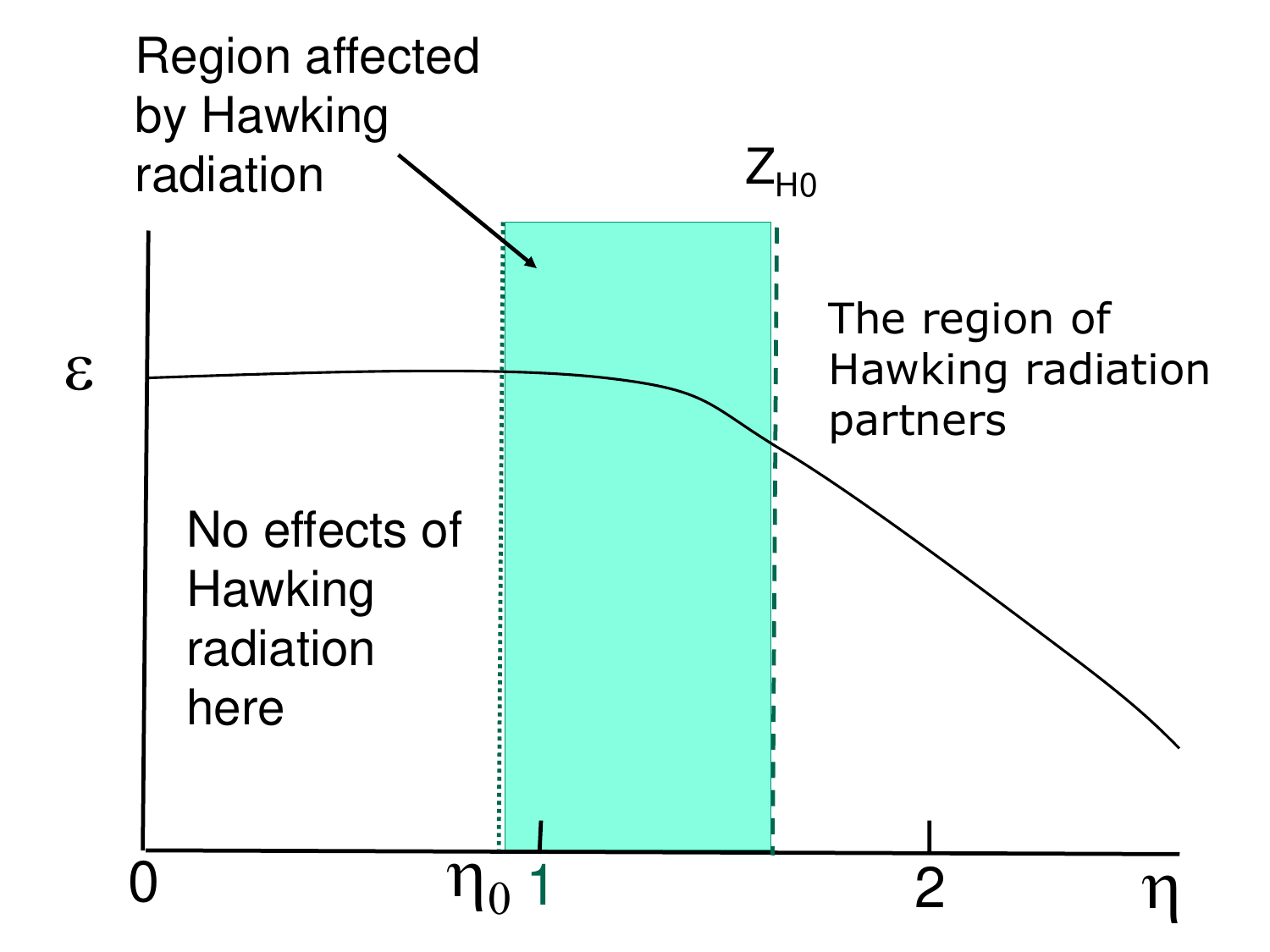}
\caption{\label{fig:fig3} Schematic representation of rapidity
regions which are affected/unaffected by Hawking radiation.}
\end{figure}

 Fig.3 shows schematic representation of rapidity regions which are 
affected/unaffected by Hawking radiation. For observer at $\eta_0$, 
the Hawking radiation will be present only for region between $\eta_0$ and 
the rapidity corresponding to the $z = z_{H0}$.  So $p_T$ distributions of
particles will be affected by Hawking radiation only in this rapidity
window. For $\eta < \eta_0$ (similarly for corresponding range of negative 
rapidities, with its own acoustic horizon), 
particle distributions will remain unaffected (depending on the 
relative speed of acoustic horizon,  for most of this region Hawking 
radiation will be infinitely red-shifted). Region with rapidities beyond the
location of acoustic horizon represent region where partner particles
(phonons) of the Hawking radiation will be found. This is the special feature
of the physics of acoustic black holes that here partner particles of 
Hawking radiation are equally accessible to experimental probes, in contrast
to actual black holes where partner particles are permanently shielded from
observations by the event horizon. In the present context, one will expect
certain specific correlations between the perturbed $p_T$ distributions 
arising from both sides of the acoustic horizon. These correlations need further
investigations (see, e.g. \cite{Steinhauer:2015saa,Liberati:2019fse,
Isoard:2019buh} in this context). Further, note that
in general one may expect a range of rapidities for the location of
observer for which a suitable acoustic horizon may exist, with its
associated regions of observable Hawking radiation. Fig.3 shows only one
such location of the observer.

  Now we discuss the implementation of the above described procedure for
identifying the acoustic horizon and the rapidity regions affected by it.
As the whole procedure crucially depends on the specific nature of the
large rapidity regions where flow does not respect Bjorken boost invariance,
we resort to results from detailed hydrodynamical simulations available in
literature.

We utilize findings of the hydrodynamic evolution of the fireball formed 
in heavy-ion collisions in ref.\cite{Bozek:2009ty}, to study the formation 
of acoustic horizon. In particular we use the results of fluid rapidity ($Y$)
vs. space time rapidity ($\eta$) at different proper times ($\tau$) during 
the evolution for $Au-Au$ collision at $\sqrt{s}=200$GeV. We have digitized
the plots of $Y(\tau,\eta) - \eta$ vs. $\eta$ from Fig.3 in 
ref.\cite{Bozek:2009ty}. The plots are given for $\tau = 2,4, $ and 6 fm/c. 
The values from the digitized plots, with interpolation, are used to get 
required values in continuous range of these variables. (Clearly, our
results are susceptible to the errors in the interpolation procedure.)

In the first step, we consider a local observer at  $\eta = \eta_0$, at 
proper time $\tau = \tau_0$. The space-time coordinates $(t_0,z_0)$ of this 
observer in the center of mass frame (which is also the lab frame), 
are given by

\begin{equation}
t_0 = \tau_0 {\rm{cosh}}(\eta_0), z_0 = \tau_0 {\rm{sinh}}(\eta_0).
\end{equation}

We calculate the fluid rapidity, $Y_0$, at the location of this observer
($t_0,z_0$), using interpolation of the $Y-\eta$ vs. $\eta$ plots from 
ref.\cite{Bozek:2009ty}. We consider the observer to be comoving with
the fluid with this fluid rapidity $Y_0$. Using $Y_0$, we compute the 
velocity and $\gamma-$factor of the fluid as well as the observer w.r.t
the lab frame as,

\begin{equation}
v_{_{z_0}}={\rm{tanh}}\left(Y_0\right), 
\gamma_{_0}={1 \over \sqrt{1 - v_{_{z_0}}^2}}.
\end{equation}

In the observer's frame of reference, at some location the fluid velocity
will be equal to the speed of sound, i.e $c_s=1/\sqrt{3}$, and will exceed
$c_s$ beyond it. This will be the location of the acoustic horizon of metric
in Eq.(1). Fluid rapidity for this 
acoustic horizon $Y_H$ will be given by

\begin{equation}
Y_H = Y_0 + Y_s,~{\rm{with}}~Y_s = 
{1\over 2}{\rm{log}}\left({1+c_s \over 1-c_s}\right) 
\end{equation}
where $Y_s$ is the fluid rapidity corresponding to fluid flowing with
the speed of sound. 
Given $Y_{H}$, we obtain the space-time rapidity and the proper time 
through the interpolation of $Y-\eta$ vs. $\eta$ plots. 
Note that for a given fluid 
rapidity, there is a range of allowed values of ($\eta,\tau$). For each of 
pair of ($\eta,\tau$), we compute the lab frame space-time coordinates of 
the acoustic horizon as,

\begin{equation}
	t = \tau {\rm cosh}(\eta),~ z=\tau {\rm sinh}(\eta),
\end{equation}
and the space-time intervals,
\begin{equation}
\Delta t= t-t_0,~\Delta z= z-z_0.
\end{equation}
Subsequently these intervals are Lorentz transformed to the 
observer's frame as,

\begin{equation}
	\Delta t_{obs}=\gamma_{_0}\left(\Delta t - v_{z_0} \Delta z \right),~
	\Delta z_{obs}=\gamma_{_0}\left(\Delta z - v_{z_0}\Delta t \right).
	\label{eqn5}
\end{equation}
Among the allowed values of ($\eta, \tau$) and corresponding ($t,z$), we 
select ($\eta_h,\tau_h$) and  ($t_h,z_h$) for which $|\Delta t_{obs}|$ is the 
lowest. Note that ($\Delta z_{obs}, \Delta t_{obs}$) are the space-time 
coordinates of the acoustic horizon in the observer frame. As the location
of acoustic horizon should be recorded at the same time as the clock
at the origin of the observer's frame, ideally $\Delta t_{obs}$ should be
zero. However, due to the fact that $Y-\eta$ vs $\eta$ data from
ref.\cite{Bozek:2009ty} is discretized, and interpolation being used has 
its own associated numerical errors, we
work with the lowest value found for $\Delta t_{obs}$. 
This is one of the sources for various errors in the calculations.

 In order to calculate the recession velocity of the acoustic horizon for
this observer, we need to calculate the horizon location at slightly later 
time. We thus consider a new proper time $\tau^\prime_0 = \tau_0 +
\delta \tau$, where $\delta \tau$ is a small interval (typically we take it
varying from 0.01 to 0.1 fm/c). During this time interval, the observer,
comoving with the fluid, would have moved from $z_0$ to a new location
$z^\prime_0$ (lab frame coordinate), with

\begin{equation}
z^\prime_0=z_0 + v_{z0}~\delta \tau.
\end{equation}
(Note $\delta \tau = \delta t$ for center of mass reference frame at
$z = 0$.) The space-time rapidity and time coordinate corresponding to this
new location of the observer are given by, 

\begin{equation}
\eta^\prime_0={\rm sinh}^{-1}(z^\prime_0/\tau^\prime_0),
t^\prime_0=\tau^\prime_0 {\rm cosh}(\eta^\prime_0).
\end{equation}
Note that we take small values of $\delta \tau$, hence the acceleration of 
the fluid with respect to the observer is neglected (another possible source 
of error).  Given ($\eta^\prime_0,\tau^\prime_0$) we obtain the fluid 
rapidity, $Y^\prime_0$, at the new location of the observer.  Note that, the 
fluid is moving with respect to the observer's new location with 
$v_{obs} = {\rm tanh}(Y^\prime_0-Y_0)$. Nonetheless, for small $\delta \tau$, 
$v_{obs}$ is found to be negligibly small. So to a good approximation, the 
observer and the fluid are co-moving also at this new location.

The new acoustic horizon has fluid rapidity $Y_H^\prime = Y^\prime_0 + Y_s$. 
Repeating above calculations, we calculate the space-time
rapidity and proper time, ($\eta^\prime_h,\tau^\prime_h$) and the space-time 
coordinates ($t^\prime_h,z^\prime_h$). We then obtain the space-time 
coordinates of the acoustic horizon in the observers frame, i.e,
\begin{equation}
\Delta t^\prime_{obs}=\gamma_{_0}\left(\Delta t^\prime - 
v_{z_0}\Delta z^\prime\right),~
\Delta z^\prime_{obs}=\gamma_{_0}\left(\Delta z^\prime - 
v_{z_0}\Delta t^\prime\right),
	\label{eqn6}
\end{equation}
where, $\Delta t^\prime = (t^\prime_{h}-t^\prime_0)$ and  
$\Delta z^\prime=(z^\prime_{h}-z^\prime_0)$. Again, from allowed values 
($\eta^\prime_h,\tau^\prime_h$), and corresponding ($t^\prime_h,z^\prime_h$),
we choose values for which $|\Delta t^\prime_{obs}|$ is lowest. 

In the lab frame, during the interval $\delta \tau$, the acoustic horizon 
has moved by $(z^\prime_h-z_h)$.  For the observer, during the interval,

\begin{equation}
\delta t_{obs} = {\delta \tau \over \gamma_{_0}}
\end{equation}
the horizon has moved by, $\delta z = \Delta z^\prime_{obs} - \Delta z_{obs}$. 
Thus, the recession velocity of the acoustic horizon with respect to 
the observer is,

\begin{equation}
	v_{hr}={\delta z \over \delta t_{obs}}.
\end{equation}
The red-shift factor of the Hawking radiation is then given by the  $\gamma-$
factor corresponding to this recession velocity,

\begin{equation}
	\gamma_{hr}=1/\sqrt{1-(v_{hr}/c_s)^2}. 
	\label{r:gf}
\end{equation}

To compute the temperature of the Hawking radiation, we consider fluid close 
to the acoustic horizon, but subsonic.  For simplicity we consider the fluid 
velocity at a point where $v_{\delta}=c_s - \delta v$, with $\delta v=0.05$. 
The fluid rapidity for this flow velocity is given by,

\begin{equation}
Y_{\delta}= Y_0 + {1 \over 2} {\rm{log}} \left({1 + v_{\delta} 
\over 1 - v_{\delta}}\right).
\end{equation}

From $Y_{\delta}$, following above steps, we compute the corresponding 
space-time
coordinate ($\Delta t_{\delta},\Delta z_{\delta}$) of the fluid element in the 
observer's frame. The gradient in the velocity profile close to the 
acoustic horizon is therefore given by (in MeV),

\begin{equation}
{dv \over dz} = 200{\delta v \over (\Delta z_{obs} - \Delta z_{\delta})},
\end{equation}
The expression for the Hawking temperature is given in Eqn.(2). Incorporating
the red-shift factor of the receding horizon (Eqn.\ref{r:gf}), the value
of Hawking radiation seen by the observer is given by (in MeV, with $z$ values
calculated in fm),

\begin{equation}
	T_{HW} = {1 \over (2\pi\gamma_{hr})}{dv \over dz},
\end{equation}

In the following we give results of our calculations in Table I and Table II. 
We consider several (sample) values of observer location $\eta_0$, and 
calculate properties of Hawking radiation from the corresponding acoustic 
horizon.  $\tau_0$ is the proper time for the observer at which Hawking 
radiation is 
calculated, $T^0_H$ represents Hawking temperature of the black hole if it 
was static, $T^R_H$ gives the Hawking temperature of the receding black 
hole incorporating the Doppler red-shift of the radiation, $f_R$ is the 
red-shift factor, and $v_H/c_s$ gives the ratio of the recession velocity 
of the black hole (w.r.t. the observer at $\eta_0$) and the sound velocity 
$c_s$.
 
 Table I gives results for a test case of Bjorken flow with longitudinal 
boost invariance by setting fluid rapidity $Y_{fl}$ equal to space-time 
rapidity $\eta$. As explained above, in this case we expect that for 
observer at any value of $\eta$, the acoustic
horizon will recede with the speed of sound. 

\begin{table}[h]
\begin{center}
\caption{Properties of acoustic black hole for Bjorken Flow with 
longitudinal boost invariance}
\begin{tabular}{||cccccc||}
\hline
$\eta_0$ & $\tau_0$ & $T^0_H$ & $T^R_H$ & $f_R$ &   $v_H/c_s$ \\ 
\hline
0.0      & 2.0      & 15.88      &  0      & $\infty$  & 1.03\\
1.0      & 2.0      & 15.88      &  0      & $\infty$  & 1.03 \\
1.5      & 2.0      & 15.88      &  0      & $\infty$  & 1.03 \\
2.0      & 2.0      & 15.88      &  0      & $\infty$  & 1.03 \\
\hline
\end{tabular}
\end{center}
\end{table}

The results in Table I show that, indeed, for Bjorken flow, Hawking
radiation is infinitely red-shifted irrespective of the location of
the observer. The recession velocity is found to be about $1.03 c_s$.
We mention that for this case, and more so for results in table II below,
there are errors in calculations arising from discretization of
various space-time intervals, and using a sequence of Lorentz transformations
(as explained above), especially at large rapidities.

 Table II presents the case of physical interest with fluid flow taken
from hydrodynamical simulations in ref.\cite{Bozek:2009ty}. 
As we mentioned above, we have used particular case of results in 
Fig.3 of ref.\cite{Bozek:2009ty} which
gives values of $Y(\tau,\eta) - \eta$ for a range of values of $\eta$. We
especially note for this case of collision with $\sqrt{s} = $ 200 GeV, 
$Y(\tau,\eta)$ differs from $\eta$ for every non-zero value of $\eta$, 
meaning that the flow deviates from Bjorken flow at every non-zero value of 
$\eta$. Thus, there is no reason to expect that acoustic
horizon will recede with speed of sound even for observer at $\eta = 0$.
This is what we see in table II where we find finite value of red-shift
for the Hawking radiation for different observer locations. Importantly,
we note that as observer's location shifts to larger values of $\eta$,
red-shift factor decreases with the value of $v_H/c_s$ decreasing form 0.91 to
0.73 as $\eta$ changes from 0 to 3. This behavior is consistent with the
main arguments of our calculations. We thus expect that for LHC with highest 
possible collision energies, we will find $v_H$ approaching $c_s$ with
infinite red-shift for small values of rapidity, and finite red-shifts for
larger values of rapidities. 
 
Note that we have considered small values of $\tau_0$. It is important to 
focus on early time $\tau_0$ because for later times the 1-d fluid expansion 
will not be a good approximation.
As the effects of Hawking radiation imply rapidity mixing of transverse momenta
of particles, even if it is set-in at very early stage, its effects should
persist for late stages also, all the way up to hadronization and freezing,
especially in certain pictures of hadronization like coalescence model
where final particle momenta basically result from constituent momenta.

\begin{table}[h]
\begin{center}
\caption{Properties of acoustic black hole for non-boost invariant 
flow from hydro simulation in ref.\cite{Bozek:2009ty}}
\begin{tabular}{||cccccc||}
\hline
\hline
$\eta_0$ & $\tau_0$ & $T^0_H$ & $T^R_H$ & $f_R$ &   $v_H/c_s$ \\ 
\hline
0.0      & 2.0      & 16.07     & 6.68   & 2.41     & 0.91\\
1.0      & 2.0      & 16.26     & 5.09   & 3.19     & 0.95\\
1.5      & 2.0      & 14.53      & 7.71       & 1.89  & 0.85\\
2.0      & 2.0      & 14.90      & 6.71       & 2.22  & 0.89\\
3.0      & 2.0      & 12.83      & 8.73       & 1.47  & 0.73\\
\hline
\end{tabular}
\end{center}
\end{table}

 In conclusion, we have presented calculations to show that even at
ultra-relativistic heavy-ion collisions, such as at LHC, with 
boost-invariant early flow for extended regions of rapidity, it is
possible to get observable effects of Hawking radiation from acoustic
horizon. In fact, in this case one expects a non-trivial rapidity
dependence for the effects with Hawking radiation affecting $p_T$
distributions of particles only at non-zero rapidities, for a finite
rapidity window. In particular, $p_T$ distributions at central rapidity
should remain unaffected. Though the steps of calculations presented here
remain valid for any general flow pattern, and in particular for the 
the hydrodynamical simulations presented in ref.\cite{Bozek:2009ty} for
$\sqrt{s}$ = 200 GeV where there is no extended rapidity window of
boost-invariant flow. 

 The Hawking radiation temperature, with the red-shift factor is found to 
 vary from 5 MeV to about 9 MeV (though without red-shift factor this
 temperature can be about 16 MeV). These values are similar to
 the values found for relatively low energy collisions which some of
 us had earlier considered in ref.\cite{Das:2020zah}. In all these cases, the
 Hawking temperature is found to be much below the expected temperature of
 the plasma. However, as we mentioned above, this has no bearing on the 
 detectability of Hawking radiation because these two temperatures, 
 namely the Hawking temperature, and the plasma temperature, relate to 
 completely different
 physical quantities. As the relevant scalar field here is fluctuations 
of fluid velocity potential, Hawking radiation will consist of quanta of 
this scalar field. This translates to perturbations in the flow,
hence in the rapidity of particles at each space-time rapidity, depending
on the strength of Hawking radiation at the respective points. This
will lead to mixing of particle distributions of nearby rapidities. The
value of Hawking temperature, and its flux, will determine the wavelength, 
and the magnitude of this rapidity mixing of $p_T$ distributions, while
the plasma temperature will continue to determine the $p_T$ distributions
of particles. The signal of the Hawking radiation will therefore be in
terms of anomalous fluctuations in momentum distributions as function
of rapidity in  different rapidity windows, as we discussed above.
The detailed nature of the signal remains to be worked out and we hope
to present it in a future work.\\

We are  very grateful to Oindrila Ganguly, Shreyansh Dave, and Saumia P.S.
for very useful discussions.



\begin{thebibliography}{99}
\bibitem{Unruh:1980cg}
W.~G.~Unruh,
Phys. Rev. Lett. \textbf{46}, 1351-1353 (1981)
doi:10.1103/PhysRevLett.46.1351

\bibitem{Visser:1997ux}
M.~Visser,
Class. Quant. Grav. \textbf{15}, 1767-1791 (1998)
doi:10.1088/0264-9381/15/6/024
[arXiv:gr-qc/9712010 [gr-qc]].

\bibitem{Steinhauer:2015ava}
J.~Steinhauer,
Phys. Rev. D \textbf{92}, no.2, 024043 (2015)
doi:10.1103/PhysRevD.92.024043
[arXiv:1504.06583 [gr-qc]].

\bibitem{Steinhauer:2015saa}
J.~Steinhauer,
Nature Phys. \textbf{12}, 959 (2016)
doi:10.1038/nphys3863
[arXiv:1510.00621 [gr-qc]].

\bibitem{MunozdeNova:2018fxv}
J.~R.~Mu\~noz de Nova, K.~Golubkov, V.~I.~Kolobov and J.~Steinhauer,
Nature \textbf{569}, 688-691 (2019)
doi:10.1038/s41586-019-1241-0
[arXiv:1809.00913 [gr-qc]].

\bibitem{Wang:2016jaj}
Y.~H.~Wang, T.~Jacobson, M.~Edwards and C.~W.~Clark,
Phys. Rev. A \textbf{96}, no.2, 023616 (2017)
doi:10.1103/PhysRevA.96.023616
[arXiv:1605.01027 [cond-mat.quant-gas]].

\bibitem{Michel:2016tog}
F.~Michel, J.~F.~Coupechoux and R.~Parentani,
Phys. Rev. D \textbf{94}, no.8, 084027 (2016)
doi:10.1103/PhysRevD.94.084027
[arXiv:1605.09752 [cond-mat.quant-gas]].

\bibitem{Grisins:2016gru}
P.~Grisins, H.~S.~Nguyen, J.~Bloch, A.~Amo and I.~Carusotto,
Phys. Rev. B \textbf{94}, no.14, 144518 (2016)
doi:10.1103/PhysRevB.94.144518
[arXiv:1606.02277 [cond-mat.quant-gas]].


\bibitem{Liberati:2019fse}
S.~Liberati, G.~Tricella and A.~Trombettoni,
Entropy \textbf{21}, no.10, 940 (2019)
doi:10.3390/e21100940
[arXiv:1908.01036 [gr-qc]].

\bibitem{Isoard:2019buh}
M.~Isoard and N.~Pavloff,
Phys. Rev. Lett. \textbf{124}, no.6, 060401 (2020)
doi:10.1103/PhysRevLett.124.060401
[arXiv:1909.02509 [cond-mat.quant-gas]].


\bibitem{Jacquet:2021scv}
M.~J.~Jacquet, L.~Giacomelli, Q.~Valnais, M.~Joly, F.~Claude, E.~Giacobino, Q.~Glorieux, I.~Carusotto and A.~Bramati,
Phys. Rev. Lett. \textbf{130}, no.11, 111501 (2023)
doi:10.1103/PhysRevLett.130.111501
[arXiv:2110.14452 [gr-qc]].

\bibitem{Syu:2022ajm}
W.~C.~Syu and D.~S.~Lee,
Phys. Rev. D \textbf{107}, no.8, 084049 (2023)
doi:10.1103/PhysRevD.107.084049
[arXiv:2212.06063 [gr-qc]].

\bibitem{Tian:2018srd}
Z.~Tian and J.~Du,
Eur. Phys. J. C \textbf{79}, no.12, 994 (2019)
doi:10.1140/epjc/s10052-019-7514-9
[arXiv:1808.03125 [quant-ph]].

\bibitem{Ganguly:2019mza}
O.~Ganguly,
[arXiv:1907.01905 [gr-qc]].

\bibitem{Kolobov:2019qfs}
V.~I.~Kolobov, K.~Golubkov, J.~R.~Mu\~noz de Nova and J.~Steinhauer,
Nature Phys. \textbf{17}, no.3, 362-367 (2021)
doi:10.1038/s41567-020-01076-0
[arXiv:1910.09363 [gr-qc]].

\bibitem{Barcelo:2005fc}
C.~Barcelo, S.~Liberati and M.~Visser,
Living Rev. Rel. \textbf{8}, 12 (2005)
doi:10.12942/lrr-2005-12
[arXiv:gr-qc/0505065 [gr-qc]].

\bibitem{Dave:2022wgk}
S.~S.~Dave, O.~Ganguly, S.~P.~S. and A.~M.~Srivastava,
EPL \textbf{139}, no.6, 60003 (2022)
doi:10.1209/0295-5075/ac8d71
[arXiv:2208.08079 [gr-qc]].


\bibitem{Das:2020zah}
A.~Das, S.~S.~Dave, O.~Ganguly and A.~M.~Srivastava,
Phys. Lett. B \textbf{817}, 136294 (2021)
doi:10.1016/j.physletb.2021.136294
[arXiv:2006.15912 [gr-qc]].


\bibitem{Nielsen:2008cr}
A.~B.~Nielsen,
Gen. Rel. Grav. \textbf{41} (2009), 1539-1584
doi:10.1007/s10714-008-0739-9
[arXiv:0809.3850 [hep-th]].



\bibitem{Nielsen:2005af}
A.~B.~Nielsen and M.~Visser,
Class. Quant. Grav. \textbf{23} (2006), 4637-4658
doi:10.1088/0264-9381/23/14/006
[arXiv:gr-qc/0510083 [gr-qc]].



\bibitem{Bozek:2009ty}
P.~Bozek and I.~Wyskiel,
Phys. Rev. C \textbf{79}, 044916 (2009)
doi:10.1103/PhysRevC.79.044916
[arXiv:0902.4121 [nucl-th]].

\end{thebibliography}

\end{document}